\documentclass[preprints,communication,accept,moreauthors,pdflatex]{Definitions/mdpi} 

\firstpage{1} 
\pubvolume{xx}
\issuenum{1}
\articlenumber{5}
\pubyear{2019}
\copyrightyear{2019}
\history{}
\usepackage{amsmath,amsfonts,amssymb}
\usepackage{bm}% bold math
\def\be{\begin{equation}}
\def\ee{\end{equation}}
\def\onehalf{{\textstyle{\frac{1}{2}}}}
\def\Aw{{\stackrel{\mbox{\tiny$\bullet$}}{\omega}}{}}
\def\omegaw{{\stackrel{\mbox{\tiny$\bullet$}}{\omega}}{}}
\def\Tw{{\stackrel{\mbox{\tiny$~\bullet$}}{T}}{}}
\def\wbol{{\stackrel{\mbox{\tiny$\circ$}}{\omega}}{}}

\Title{Gauge Structure of Teleparallel Gravity$^{\mbox{\small$\ddag$}}$}
\Author{Jos\'e G. Pereira$^{1}$, Yuri N. Obukhov$^{2}$}

\address{%
$^{1}$\quad Universidade Estadual Paulista (UNESP), Instituto de F\'{\i}sica Te\'orica, 
05409-011 S\~ao Paulo, Brazil; jg.pereira@unesp.br\\
$^{2}$\quad Russian Academy of Sciences, Nuclear Safety Institute (IBRAE),
B. Tulskaya 52, 115191 Moscow, Russia; obukhov@ibrae.ac.ru}
\secondnote {Contribution to the proceedings of the workshop {\em Teleparallel Universes in Salamanca}, Salamanca, Spain,  26-28 November 2018.}

\abstract{During the conference {\em Teleparallel Universes in Salamanca}, we became aware of a recent paper [M. Fontanini, E. Huguet, and M. Le Delliou, {\em Phys. Rev. D} {\bf 2019}, {99}, 064006] in which some criticisms on the interpretation of teleparallel gravity as a gauge theory for the translation group were put forward. This triggered a discussion about the arguments on which those criticisms were based, whose output is described in the present paper. The main conclusion is that, to a great extent, those arguments are incorrect, and lack mathematical and physical support.}

\begin{document}

\section{Lessons on Bundles and Gauge Theories}
\label{S1}

\subsection{The Frame Bundle}
\label{S11}

The frame bundle is a constitutive part of spacetime: it is always present as soon as spacetime\footnote{For practical purposes, we use the same notations of Ref.~\cite{TEGR}} $M$, the base space of the bundle, is assumed to be a differentiable manifold \cite{GEOlivro}. The frame bundle is usually considered to be the prototype of a principal bundle. The whole bundle space is locally the Cartesian product $M \times G$, where $G$ is the structure group of the bundle. In the most general case, $G$ is the general linear group of matrices $GL(4,{\mathbb R})$. Once spacetime $M$ is endowed with a Lorentzian metric ${\bm g}$, one can define a sub-bundle called the bundle of orthonormal frames, which consists of orthonormal bases with respect to ${\bm g}$, and with the Lorentz group $SO(1,3)$ as the structure group \cite{HE}. Technically, the orthonormality condition for the frames is introduced with the help of the soldering form, see \cite{Marsh} for more details. The fiber over a point $x \in M$ is the set of all frames (ordered bases) for the tangent space $T_x M$. These fibers can always be glued together, through a local Cartesian product, in such a way to yield a principal orthonormal $SO(1,3)$ bundle over $M$.

In special relativity, the bundle of orthonormal frames, whose structure group is the Lorentz group, deals essentially with inertial effects. Different classes of frames are obtained by performing {\it local} Lorentz transformations. Within each class, the infinitely many frames are related to each other by {\it global} Lorentz transformations. Of course, any gauge theory describing an interaction of Nature has its own principal bundle. For example, the electromagnetic field is described by a gauge theory for the one-dimensional unitary group $U(1)$. The principal bundle of the electromagnetic theory is consequently given locally by the Cartesian product $M \times U(1)$. Since the group manifold of $U(1)$ is the sphere $S^1$, this bundle is locally homeomorphic to $M \times S^1$. The role of the frame bundle in this case is just to describe how the electromagnetic field equation changes when seen from different classes of frames.

\subsection{The Gauge-Theoretic Bundle Framework of Teleparallel Gravity}
\label{S12}

Let us consider now teleparallel gravity, a gauge theory for the translation group $T_4$ \cite{Cho}. As a gauge theory, and similarly to electromagnetism, its principal bundle is given locally by the Cartesian product $M \times T_4$. Considering furthermore that the group manifold of $T_4$ is the Minkowski space ${\mathbb R_4}$, this bundle is locally homeomorphic to the tangent bundle $M \times {\mathbb R_4}$. A crucial point of this construction is to note that there are two different notions of Minkowski space involved. The first notion is the usual one, which appears as the tangent space to spacetime, and on which the Lorentz transformations of the frame bundle take place. This Minkowski space is spanned by the tangent vectors at a point of spacetime, and is of course a vector space. It is not, however, the Minkowski space appearing as fiber in the tangent bundle of teleparallel gravity. In fact, instead of being spanned by the tangent vectors at a point of spacetime, this Minkowski space is constructed as the point set of the translation group $T_4$. As a point set, such a Minkowski is not a vector space, but an affine generalization of Minkowski \cite{Minko}. Gauge translations take place in this affine Minkowski space.\footnote{In differential geometry, this kind of space is sometimes called a $G$-torsor, where $G$ is the group whose point set is the space itself \cite{torsor1}. For an intuitive, physically-motivated description of the concept of torsor as a principal  homogeneous space (i.e. a space that carries a free and transitive action of the structure group), see Ref.~\cite{torsor2}.}

It should be noted that this is not an inherent property of teleparallel gravity, but a general feature of gauge theories. For example, consider the electromagnetic theory, a gauge theory for the one-dimensional unitary group $U(1)$. The internal space of this theory, given by the point set of the group $U(1)$, is an affine version of the sphere $S^1$. The electromagnetic gauge transformations take place in this affine space. As a matter of fact, gauge transformations always take place in affine spaces, where no origin is specified~\cite{KN}. 

With the above proviso, teleparallel gravity is easily found to admit a consistent formulation in terms of connections on principal fiber bundles. The main geometrical structure is the principal fiber bundle $P(M,G)$ over the four-dimensional spacetime manifold $M$ with the structure group $G = T_4$ of translations. Most conveniently, the latter can be realized as the translation group on the tangent space $T_xM$, viewed as an affine space $A_xM$. Describing an element $v\in A_xM$ as a 5-vector
\[
v = \left(\begin{array}{c}v^a \\ 1\end{array}\right),
\]
one can describe a translation $g \in T_4$ as a $5\times 5$ matrix
\[
g = \left(\begin{array}{cc}{\mathbb I} & \chi^a \\ 0 & 1\end{array}\right).
\]
With a $4\times 4$ unit matrix ${\mathbb I}$ and four parameters $\chi^a$, the translation on $A_xM$ is then represented by
\be
v \rightarrow v' = gv = \left(\begin{array}{cc}{\mathbb I} & \chi^a \\ 0 & 1\end{array}\right)\left(\begin{array}{c}v^a \\ 1\end{array}\right) = \left(\begin{array}{c}v^a + \chi^a \\ 1\end{array}\right).
\ee

One can view $\left\{\chi^a\right\}$ as the local coordinates on $G$, and then together with the local coordinates $\left\{x^\mu\right\}$ on $M$, we have the local coordinates $\left\{x^\mu, \chi^a\right\}$ on the principal bundle $P(M,G)$. Although the bundle formalism is well developed for the description of the global features, we will confine our attention just to the local coordinate systems, which is sufficient for all practical purposes to understand the teleparallel gravity as a gauge theory of the group of translations. Denoting the points of the bundle $u\in P$ and base space $x\in M$, the canonical projection
\[
\pi: P\rightarrow M, \quad \pi(u) = x
\]
explicitly reads as $\pi(x^\mu, \chi^a) = x^\mu$. The local section
\[
\sigma: M\rightarrow P,\quad \sigma(x) = u,
\]
is obviously described as $\sigma = (x^\mu, \chi^a(x))$. Given the local coordinates on $M$ and $P$, we have the natural frames on the corresponding tangent spaces $\partial_\mu \in T_xM$ and $(\partial_\mu, \partial_a) \in T_uP$, where we use the condensed notation $\partial_\mu = {\partial}/{\partial x^\mu}$ and $\partial_a = {\partial}/{\partial \chi^a}$. The canonical projection $\pi: P\rightarrow M$ and the local section $\sigma: M\rightarrow P$ define the corresponding differential (``push-forward'') maps of the tangent spaces
\be
\pi_*: T_uP\rightarrow T_xM \quad {\rm and} \quad \sigma_*: T_xM\rightarrow T_uP \, ,
\ee
which in the local coordinates are evidently given by
\begin{equation}\label{dpisi}
\pi_*(\partial_\mu, \partial_a) = \partial_\mu \quad {\rm and} \quad
\sigma_*(\partial_\mu) = \partial_\mu + {\frac {\partial \chi^a}{\partial x^\mu}}\,\partial_a\,. 
\end{equation}

A connection on the principal fiber bundle determines the decomposition of the tangent space $T_uP$ into the sum of the horizontal and vertical subspaces. The latter is conveniently spanned by the fundamental vector fields on $P$, which are determined by the right action of the structure group $G$ on $P$, so that any Lie algebra element $\xi\in {\cal G}$ induces a vector field (which by construction is therefore vertical) $\xi^*$. For $G = T_4$, in the local coordinates for the translations $\xi^a\in {\cal G}$ we have the fundamental vector field $\xi^* = \xi^a\partial_a$.

The connection is introduced by the 1-form $\omega^a$ on $P$ with the values in the Lie algebra of the group of translations, such that its value on the fundamental vector field is $\omega^a(\xi^*) = \xi^a$. In the local coordinates $(x^\mu, \chi^a)$, the connection 1-form with the required property is given by
\begin{equation}
\omega^a = B^a{}_\mu\,dx^\mu + d\chi^a.\label{conn}
\end{equation}
For every vector $v\in T_xM$, there exists a unique horizontal lift $\widetilde{v}$, i.e., a vector in the horizontal subspace $\widetilde{v}\in T_uP$ such that under the canonical projection $\pi_*(\widetilde{v}) = v$. Explicitly, a horizontal lift for the coordinate frame is
\begin{equation}
\widetilde{\partial}{}_\mu = \partial_\mu - B^a{}_\mu\,\partial_a.\label{lift}
\end{equation}
One can check that $\omega^a(\widetilde{\partial}{}_\mu) = 0$, hence the connection 1-form is vertical, as it should. As a result, we can write the decomposition of an arbitrary vector $T_uP \ni V = \alpha^\mu\partial_\mu + \beta^a\partial_a$ into the horizontal and vertical parts as
\begin{equation}
V = \alpha^\mu\left(\partial_\mu - B^a{}_\mu\partial_a\right) + \left(\beta^a + B^a{}_\mu\alpha^\mu\right)\partial_a.\label{hor}
\end{equation}

Given a local section, $\sigma: M\rightarrow P$, we obtain the connection 1-form on spacetime manifold $M$ via the pull-back map
\begin{equation}
h^a := \sigma^*\omega^a = \left(B^a{}_\mu + \partial_\mu\chi^a\right)dx^\mu.\label{ha} 
\end{equation}
To put it differently, we obtain the components of the connection form on the base space from (\ref{dpisi}):
\be
\omega^a(\sigma_*\partial_\mu) = B^a{}_\mu + \partial_\mu\chi^a =: h^a{}_\mu \, .
\label{coframe}
\ee
Since the coframe $h^a$ is gauge invariant, it cannot be gauged away \cite{Handbook}. Then, following the usual gauge-theoretic lines, we obtain a physical interpretation of the fiber bundle scheme by identifying the connection 1-form $B^a$ with the translational gauge potential of the gravitational field, in which case the coframe $h^a$ is non-trivial and the corresponding translational gauge field strength 2-form $\Tw^a = dh^a \equiv {\frac 12}\,\Tw^a{}_{\mu\nu}\,dx^\mu\wedge dx^\nu$, with
\be
\Tw^a{}_{\mu\nu} = \partial_\mu h^a{}_\nu - \partial_\nu h^a{}_\mu \, ,
\label{Torsionh}
\ee
is interpreted as the torsion for the coframe $h^a$ \cite{TeleBook}. One should impose a natural condition that the latter is non-degenerate when we deal with classical gravity. In the context of quantum gravity, however, one can allow also for a degenerate coframe, which gives rise to the so-called Carroll geometry on the spacetime manifold.

This completes the construction of kinematics of the teleparallel gravity as a gauge theory of the translation group. The ensuing dynamics can be obtained by following the usual steps of the gauge paradigm \cite{Hodge}.
Most straightforwardly, the dynamics of teleparallel gravity can be obtained in the framework of the premetric approach. Technically, this amounts to the construction of the Yang-Mills type Lagrangian \cite{epjc}
\begin{equation}\label{L}
L = -\,{\frac 18}\,\chi^{\mu\nu}{}_a{}^{\rho\sigma}{}_b\,\Tw^a{}_{\mu\nu}\Tw^b{}_{\rho\sigma}\,\eta,
\end{equation}
where $\chi^{\mu\nu}{}_a{}^{\rho\sigma}{}_b$ is the corresponding constitutive tensor, and $\eta$ is the volume 4-form. Since the tetrad is gauge invariant, the constitutive tensor $\chi^{\mu\nu}{}_a{}^{\rho\sigma}{}_b$ and the volume 4-form $\eta$ are both gauge invariant, which assures the gauge invariance of the gravitational Lagrangian (\ref{L}).

%%%%%%%%%%%%%%%%%%%%%%%%%%%%%%%%%%%%%%%%%%%
\section{Lessons on Spin Connections and Frames}
\label{S2}

The role played by the Lorentz (or spin) connection in teleparallel gravity seems to be a constant source of serious misunderstandings in the literature. Strictly speaking, Lorentz connections are alien objects in the field-theoretic fiber bundle formulation of the translational gauge theory, in the sense that they have nothing to do with gravitation. Nevertheless, they are behind all relativistic theories, where they describe inertial effects present in a given class of frames. Of course, they are also behind teleparallel gravity, and can accordingly be consistently introduced into the theory, where they play the same role played in all other relativistic theories. 

%%%%%%%%%%%%%%%%%%%%%%%%%%%%%%%%%%%%%%%%%%%
\subsection{Spin Connection in Special Relativity}
\label{S21}

Let us consider a relativistic free particle. In the class of inertial frames, denoted here by $e'^a{}_\mu$, its equation of motion has the simple form
\begin{equation}
\frac{d u'^a}{d\tau} = 0,
\label{EM25}
\end{equation}
with $u'^a$ the anholonomic particle four--velocity, and $d\tau$ the Minkow\-ski interval. Since it is written in a specific class of frames, equation \eqref{EM25} is not \emph{manifestly} covariant under local Lorentz transformations. This does not mean, however, that it is not covariant. In fact, in a general Lorentz rotated frame $h^a{}_\mu = \Lambda^a{}_b h'^b{}_\mu$, the equation of motion (\ref{EM25}) assumes the Lorentz covariant form
\begin{equation}
\frac{d u^a}{d\tau} + \Aw^a{}_{b \mu} \, u^b \, u^\mu = 0,
\label{anholoEM}
\end{equation}
where $u^a = \Lambda^a{}_b(x) \, u'^b$ is the Lorentz rotated four--velocity, with $u^\mu \equiv u^a h_a{}^\mu = {d x^\mu}/{d\tau}$ the corresponding spacetime holonomic four--velocity. In this expression,
\begin{equation}
\Aw^a{}_{b \mu} = \Lambda^a{}_e(x) \, \partial_\mu \Lambda_b{}^e(x)
\label{InerConn}
\end{equation}
is a Lorentz--valued (spin) connection that represents the inertial effects present in the Lorentz--rotated frame $h^a{}_\mu$. In special relativity, therefore, similarly to all other relativistic theories, the inertial spin connection (\ref{InerConn}) shows up naturally when moving through different classes of Lorentz frames. The geometrical structure in charge of describing those kinematic effects is the frame bundle, which like the inertial spin connection, has nothing to do with gravitation. In Appendix A, we show that the inertial connection (\ref{InerConn}) is just the Levi-Civita connection of flat spacetime.\footnote{The Levi-Civita connection is a special case of local Lorentz connection, which is uniquely determined by the metric-compatibility and the torsion-free conditions.}

%%%%%%%%%%%%%%%%%%%%%%%%%%%%%%%%%%%%%%%%%%%
\subsection{Spin Connection in Teleparallel Gravity}
\label{S22}

In general relativity, a theory grounded on the equivalence principle, the Levi-Civita spin connection $\wbol^a{}_{b \mu}$ represents both gravitation and inertial effects. In teleparallel gravity, on the other hand, gravitation and inertial effects are represented by different variables. In fact, whereas gravitation is represented by a translation--valued connection $\omega^a$ 1-form (\ref{conn}), inertial effects are represented by the same Lorentz connection $\Aw^a{}_{b \mu}$ appearing in special relativity. To see how such spin connection shows up in the context of teleparallel gravity, let us recall that, even though there is no inertial frames in the presence of gravitation, it is possible to define a class of frames, called {\it proper frames}, in which no inertial effects are present \cite{ObuPer}, and consequently the inertial spin connection $\Aw^a{}_{b \mu}$ vanishes globally. In this class of frames, the translational covariant derivative of a general source field $\Psi$ is written as \cite{illumi}
\begin{equation}
h_\mu \Psi = \partial_\mu \Psi + B^a{}_{\mu} \, \partial_a \Psi \, .
\label{TransCoVa}
\end{equation}
Owing to the soldering property of the tangent bundle, the above covariant derivative can be rewritten in the form $h_\mu \Psi = h^a{}_\mu  \partial_a \Psi$, with
\begin{equation}
h^a{}_\mu = \partial_\mu \chi^a + B^a{}_\mu
\label{tetrad0}
\end{equation}
the teleparallel tetrad, or coframe (see Eq.~(\ref{coframe})).

Of course, since this covariant derivative holds in the specific class of proper frames, it is not manifestly covariant under local Lorentz transformations. In order to obtain its Lorentz covariant form, one has just to perform a local Lorentz transformation
\begin{equation}
\chi^a \to \Lambda^a{}_b(x) \,\chi^b.
\end{equation}
Considering that the translational gauge potential is a Lorentz vector in the algebraic index, that is, $B^a{}_\mu ~\to~ \Lambda^a{}_b(x) \, B^b{}_\mu$, it is easy to see that in a general Lorentz frame the translational covariant derivative \eqref{TransCoVa} assumes the form
\begin{equation}\label{TransCova2}
h_\mu\Psi = \partial_\mu\Psi + \Aw^a{}_{b\mu}\chi^b\,\partial_a\Psi + B^a{}_\mu\partial_a\Psi
\end{equation}
with $\Aw^{a}{}_{b\mu}$ the same inertial Lorentz connection \eqref{InerConn} of special relativity. As in the specific case of proper frames, owing to the soldered property of the tangent bundle, the above covariant derivative can again be rewritten in the form $h_\mu \Psi = h^a{}_\mu  \partial_a \Psi$, with
\begin{equation}
h^a{}_\mu = \partial_\mu \chi^a + \Aw^a{}_{b \mu} \chi^b + B^a{}_\mu
\label{NonTriTetra}
\end{equation}
the Lorentz rotated teleparallel coframe. Notice that, to each tetrad there is a Lorentz connection $\Aw^a{}_{b \mu}$ that represents the inertial effects present in the frame represented by $h^a{}_\mu$ \cite{MarPer}. Notice also that the teleparallel torsion is defined as the covariant derivative of the coframe in the very same inertial connection appearing within the coframe:
\begin{equation}\label{tfs3}
\Tw^a{}_{\mu \nu} = \partial_\mu h^a{}_\nu - \partial_\nu h^a{}_\mu +
\Aw^a{}_{b \mu} h^b{}_\nu - \Aw^a{}_{b \nu} h^b{}_\mu \, .
\end{equation}
In the class of proper frames, where the spin connection vanishes, torsion is defined by Eq.~(\ref{Torsionh}), with the tetrad given by (\ref{coframe}).

Although ubiquitous in teleparallel gravity, the inertial spin connection $\Aw^a{}_{b\mu}$ is, as already mentioned, irrelevant for the dynamics of the gravitational field. This can be easily seen from the fact that, like any other special relativistic theory, teleparallel gravity can be equivalently described in the class of proper frames, where the inertial connection vanishes, or in a general class of frames, where the inertial connection is non-vanishing. In either case, the solution to the gravitational field equations will be the same. In fact, through a lengthy but straightforward computation, one can verify that the inertial spin connection $\Aw^a{}_{b\mu}$ enters the teleparallel gravity Lagrangian through a surface term, and consequently it does not contribute to the gravitational field equations \cite{MarPer}. This constitutes a clear evidence that, in teleparallel gravity, the Lorentz invariance is not a gauge symmetry, but just a kinematic symmetry.

%%%%%%%%%%%%%%%%%%%%%%%%%%%%%%%%%%%%%%%%%%%
\section{Discussion and Conclusions}
\label{S3}

In Ref.~\cite{TEGR}, the authors have made a series of criticisms on the translational gauge approach to teleparallel gravity. Their basic conclusion is that teleparallel gravity cannot be formulated in terms of the usual fiber bundle mathematical structure underlying a gauge theory for a symmetry group $G$. In their reasoning, however, there is a number of misunderstandings which should be clarified.

To begin with, the most important issue is perhaps the question about the principal bundle of teleparallel gravity. As discussed in Sec.~\ref{S1}, the principal bundle of teleparallel gravity, seen as a gauge model for the translation group $G = T_4$, is locally given by the Cartesian product $M \times T_4$, with $M$ the base space of the bundle, which in this case is spacetime itself. The principal bundle is naturally obtained by identifying $T_4$ with its point set, which is an affine generalization of Minkowski spacetime. For some reason, however, the authors assumed in \cite{TEGR} that the principal bundle of teleparallel gravity is the frame bundle. This is clearly not correct, and such an assumption leads indeed to consistency problems. In fact, in the last paragraph of page~5 we read: {\it One could ask if there is a possibility to associate in a natural way a principal bundle with the tangent bundle. The answer is yes but it turns out that this bundle is precisely the frame bundle, and no translations are present there}. In fact, no translations are present in the frame bundle. However, instead of merely concluding that the frame bundle cannot be identified as the principal bundle of the teleparallel gravity, the authors concluded that the gauge approach to the teleparallel gravity is inconsistent. This is of course incorrect.

Except in general relativity, whose spin connection includes both gravitation and inertial effects, in all other relativistic theories the frame bundle has to do with inertial effects only. These effects are represented by the purely inertial Lorentz connection $\Aw^a{}_{b \mu}$, which as discussed in Sec.~\ref{S2} is different for different classes of frames. These classes of frames are related to each other by local Lorentz transformations, which make up the structure group of the frame bundle. It is important to remark that the inertial spin connection has nothing to do with gravity. Its role is just to render all relativistic theories invariant under local Lorentz transformations. In this sense, the presence of the inertial spin connection $\Aw^a{}_{b \mu}$ in teleparallel theory is just to enforce the explicit local Lorentz invariance of the theory. The statement made on page 6 of \cite{TEGR} that {\it the local Lorentz invariance is not satisfied by a translational field alone}, is therefore misleading. In addition, in the first paragraph of \cite{TEGR} one reads that the {\it teleparallel gravity is obtained by choosing the Weitzenb\"ock connection, instead of the Levi-Civita connection of general relativity}. However, in the light of the discussion presented above, like in any relativistic theory, the spin connection of teleparallel gravity does not need to be chosen: it vanishes in the class of proper frames, and shows up naturally in the theory upon moving to different classes of Lorentz frames.

Another relevant point is the statement made on page 6 that {\it the Lorentz symmetry is implicitly gauged in teleparallel gravity}. This is clearly not the case. Since the corresponding spin connection is a purely inertial connection, in the teleparallel gravity the local Lorentz group is obviously a kinematic and not a dynamic (or gauge) symmetry. In other words, there is no physical gauge field corresponding to the local Lorentz group. Moreover, moving to the proper class of frames, the inertial spin connection vanishes globally. If the spin connection were a true physical field, it could never be gauged away from the theory. The argument raised in \cite{TEGR} to support a Cartan connection approach to teleparallel gravity, consequently, also turns out to be incorrect.

It should be remarked that, although the arguments to support a Cartan connection approach to teleparallel gravity do not hold, one certainly should not underestimate the usefulness of the powerful geometrical methods related to Cartan's connection when considering teleparallel gravity in the broader context of metric-affine gravity (MAG) theory \cite{MAG}. The latter has a natural interpretation as the gauge theory \cite{Reader} of the general affine group $GA(4, {\mathbb R}) = GL(4, {\mathbb R})\rtimes T_4$, the semi-direct product of the general linear group and the group of translations. The fiber bundle formulation in this case can be consistently constructed in terms of affine connections on the principal bundle of affine frames \cite{Book}. 

Summarizing, one can say that the criticisms of Ref.~\cite{TEGR} on the interpretation of teleparallel gravity as a gauge theory for the translation group are, to a great extent, incorrect. For example, it is claimed on page 6 that {\it a fiber bundle for the translations fails to be principal, except if it is trivial}. However, in the field-theoretic fiber bundle formulation of teleparallel gravity, the Minkowski spacetime that appears as fiber of the tangent bundle is not a vector space, but an affine generalization of Minkowski in which the origin is not fixed \cite{KN}. This means that the bundle of teleparallel gravity is not a vector bundle. As a consequence, it does not admit a global section, and is in general nontrivial. Contrary to the above mentioned claim, and as outlined in Sec.~\ref{S12}, teleparallel gravity can be consistently formulated in terms of the translational connection on a principal bundle \cite{TeleBook}.

\funding{JGP thanks CNPq-Brazil (Grant No. 304190/2017-9) for partial financial support. The work of YNO was partially supported by the Russian Foundation for Basic Research (Grant No. 18-02-40056-mega).}

\acknowledgments{The authors are grateful to the Organizers of the workshop {\em Teleparallel Universes in Salamanca} (26-28 November 2018) for the invitation and warm hospitality.}

%%%%%%%%%%%%%%%%%%%%%%%%%%%%%%%%%%%%%%%%%%
%% optional
\appendixtitles{yes} %Leave argument "no" if all appendix headings stay EMPTY (then no dot is printed after "Appendix A"). If the appendix sections contain a heading then change the argument to "yes".
\appendix
\section{The Flat Levi-Civita Connection}
%\unskip
%\subsection{}

As discussed in Section \ref{S21}, in the inertial frame $h'^a{}_\mu$, where no inertial effects are present, the spin connection $\omegaw'^a{}_{b \mu}$ vanishes. In the Lorentz rotated frame
\be
h^b{}_\mu = \Lambda^b{}_a(x) \, h'^a{}_\mu \, ,
\label{1}
\ee
the inertial connection is no longer vanishing, and assumes the form
\begin{equation}
\omegaw^a{}_{b \mu} = \Lambda^a{}_e(x) \, \partial_\mu \Lambda_b{}^e(x) \, .
\label{2}
\end{equation}
Multiplying both sides of Eq.~(\ref{1}) by $h'_c{}^\mu$, and using the orthogonality of the coframe, we get
\be
\Lambda^b{}_c(x) = h^b{}_\mu \, h'_c{}^\mu \, .
\label{3}
\ee
The inverse matrix is easily found to be
\be
\Lambda_c{}^b(x) = h_c{}^\mu \, h'^b{}_\mu \, .
\label{4}
\ee
Substituting (\ref{3}) and (\ref{4}) into (\ref{2}), and using the notation
\be
\omegaw^a{}_{b c} = h_c{}^\mu \omegaw^a{}_{b \mu} \, , 
\ee
a simple computation yields
\begin{equation}%
\omegaw^{a}{}_{b c} = \onehalf \left(f_{b}{}^{a}{}_{c} 
+ f_{c}{}^{a}{}_{b} - f^{a}{}_{b c} \right) 
\label{tobetaken30}
\end{equation}%
where
\begin{equation}
f^c{}_{a b} = h_a{}^{\mu} h_b{}^{\nu} (\partial_\nu
h^c{}_{\mu} - \partial_\mu h^c{}_{\nu} )
\label{fcab0}
\end{equation}
is the coefficient of anholonomy of the Lorentz rotated frame $h^a{}_\mu$. Equation (\ref{tobetaken30}) is the usual expression of the Levi-Civita connection in terms of the coefficient of anholonomy. This shows that the inertial connection (\ref{2}) and the Levi-Civita connection (\ref{tobetaken30}) for a flat spacetime are one and the same connection. It is important to remark that this result holds strictly in the flat Minkowski spacetime of special relativity. In the case of teleparallel gravity, discussed in Section~\ref{S22}, the inertial connection is no longer a Levi-Civita connection.

%\section{}
%All appendix sections must be cited in the main text. In the appendixes, Figures, Tables, etc. should be labeled starting with `A', e.g., Figure A1, Figure A2, etc. 

%%%%%%%%%%%%%%%%%%%%%%%%%%%%%%%%%%%%
\reftitle{References}

\end{document}